\documentclass[review]{elsarticle}

\usepackage{lineno,hyperref}
\modulolinenumbers[5]

\journal{Journal of \LaTeX\ Templates}

\begin{document}

\begin{frontmatter}

\title{Compositional analysis of laser produced plasma plume in front and back ablation geometries}

\author[mymainaddress1]{Alamgir Mondal}
\author[mymainaddress2]{R. K. Singh}
\author[mymainaddress1]{H. C. Joshi}


\cortext[mycorrespondingauthor]{Corresponding author}
\ead{hem@ipr.res.in}

\address[mymainaddress1]{Institute for Plasma Research, HBNI, Bhat, Gandhinagar, Gujrat 382428, India}
\address[mymainaddress2]{Institute for Plasma Research, Bhat, Gandhinagar, Gujrat 382428, India}

\begin{abstract}
Composition analysis of LiF-C thin film for neutral and ionic contributions in Front Ablation (FA) and Back Ablation (BA) geometries in vacuum and $2\times10^{-1}$ mbar argon has been done. Temporal evolution of ionic to neutral ratio and neutral abundance for both the geometries has been estimated.  For neutral abundance, two approaches viz Atomic Data and Analysis Structure (ADAS) analysis as well as integrated intensity ratio of Li I 670.1 nm and Li I 610.3 nm lines assuming LTE conditions are explored. The present attempt will be interesting from the view point of understanding the evolution of plasma composition in various geometries/configurations of laser ablation which has important implications in various applications e.g. pulsed laser deposition and laser cleaning.
\end{abstract}


\end{frontmatter}

\section{Introduction}

\begin{figure}
\begin{center}
\includegraphics[width = 6 cm]{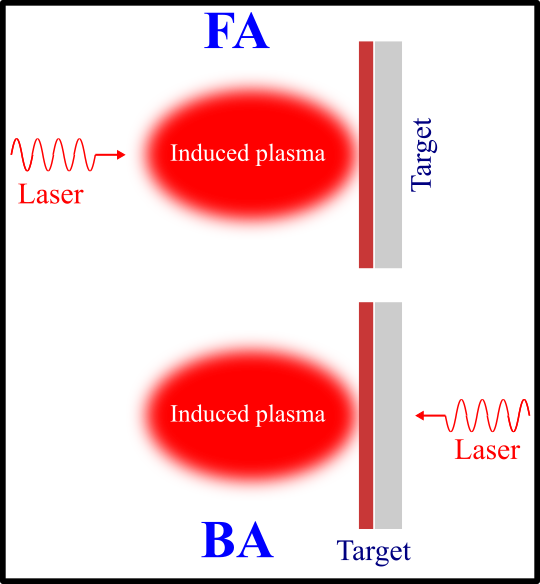}
\caption{\label{fa_ba} Schematic diagram of Front Ablation (FA) and Back Ablation (BA) geometries.}
\end{center}
\end{figure}

The properties of the laser induced plasma have been a key area of interest because of its immense potential in fundamental as well as application oriented works. It has tremendous applications viz. laser processing of materials, elemental analysis, thin film deposition, ion source generation and plasma diagnostics.\cite{cristoforetti2005libs,bohandy1986filmdepos,zepf2001ionsource,wiley} Rear surface or back ablation (BA) of thin film (Fig. \ref{fa_ba} ) deposited on a transparent substrate using laser beam is promising from the application point of view. In this case, the laser is incident from the rear side and heats the target material at illuminated region and expel the molten material in form of plasma plume along the laser direction, which is also called as laser blow-off.\cite{bohandy1986filmdepos,bkumar2014} Because of the specific nature of plasma plume, it has better prospects in the deposition of thin film, material science, micro fabrication, beam generation of energetic particles and diagnostics.\cite{bohandy1986filmdepos, wegner2018tomakdiag,fernandez2018microfab}

Besides this, front ablation (FA) of thin film (Fig. \ref{fa_ba} ) is another mechanism where the film deposited on a transparent substrate is ablated by incident laser in a conventional method which can be seen in the upper image in Fig\ref{fa_ba}. Front ablation of thin film can also facilitate range of applications such as removal of thin film, laser cleaning etc.\citep{bohandy1986filmdepos,mayer1985plasma} In view of above, a comparative study of LiF-C thin film in FA and BA geometries is an important aspect to be carried out under various experimental scenarios.

Further, compositional analysis of the plume is an important aspect for controlling the plasma parameters through optimization of the conditions in case of pulsed laser deposition and laser cleaning. Recently it has been reported that back ablation is favourable for better film deposition and cleaning of glassware.\cite{gmbilmes2018, lpabst2018} Some works were reported regarding the study of front and rear ablation in laser plasmas.\cite{escobarfaba2002} However, a proper compositional analysis was not attempted in the past. In our previous work, we reported some salient feature of plasma plume dynamics.\cite{alam2018} In the present communication we report detailed compositional analysis of the evolution of lithium plasma in back and front ablation geometries (Fig. \ref{fa_ba}) considering ion-neutral abundance ($n_i / n_a$) and neutral abundances for BA and FA ($n_{BA}/n_{FA}$) by exploiting optical emission spectroscopy (OES). We have considered two lines of neutral lithium viz. Li I 670.8 nm (resonance line) and Li I 610.3 nm (non-resonance line) for analysis. A schematic energy level diagram depicting these two lines is shown in figure \ref{E_label}. As can be seen that 670.8 nm line is a resonance lines terminating in the ground state whereas 610.3 nm line terminates at 1.85 eV.

\section{Experimental set-up}
The experimental set up is the same which was used in our earlier studies.\cite{alam2018} Briefly, an Nd:YAG laser having fundamental wavelength of 1064 nm (pulse width $\sim$8 ns at FWHM) is used to create plasma plume of LiF-C thin film target in two different geometries i.e. FA and BA. LiF-C thin film (0.5 $\mu$m carbon and 0.05 $\mu$m LiF) deposited on a transparent quartz plate of 2 mm thick is used as target. The thin film coated target plate ( 60 mm $\times$ 60 mm ) is placed in a holding arrangement normal to the laser beam. The whole arrangement is kept at the center of a cylindrical stainless steel vacuum chamber. The target holder is fixed with a base which is connected to a motor and a position micro-controller. The experiment has been done in single shot mode and the required laser fluence onto the target surface has been achieved by controlling the laser operating parameters.

Optical emission spectroscopy (OES) has been used to calculate the plasma electron temperature and density. A custom made collimator (Andor Collimator, model- ME-OPT-0007) connected to a spectrometer (0.5 m, Acton Advanced SP2500A, Princeton Instruments) through optical fibre collects the emission (up to 4 mm from target surface) of the transient plasma. The spectrometer output is coupled with an ICCD and it has $\sim$ 0.08 nm resolution. A micro-controller based timing generator having $<$ 1 ns timing jitter synchronises the spectrograph, ICCD camera and laser. Emission is collected from the same location of the plume after 2 mm distance from the target surface for both BA and FA geometries and experimental conditions are also maintained same. Moreover, the spectrometer detector response is uniform for the spectral lines taken in the study.

\section{Results and Discussion}

\begin{figure}
\begin{center}
\includegraphics[width = 6 cm]{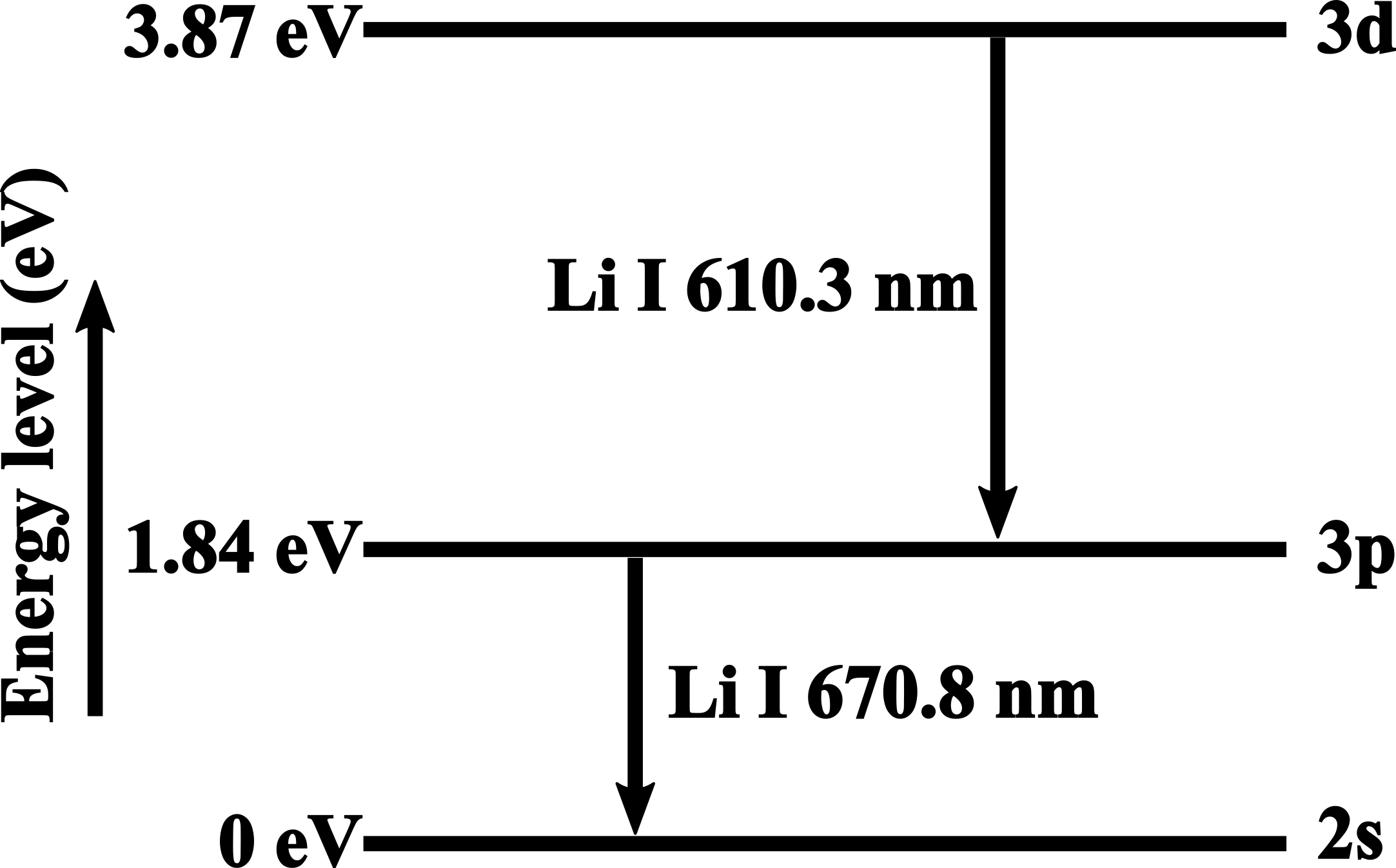}
\caption{\label{E_label} Energy level diagram showing Li I 610.3 nm (non-resonance) and Li I 670.8 (resonance) nm transitions.}
\includegraphics[width = 6 cm]{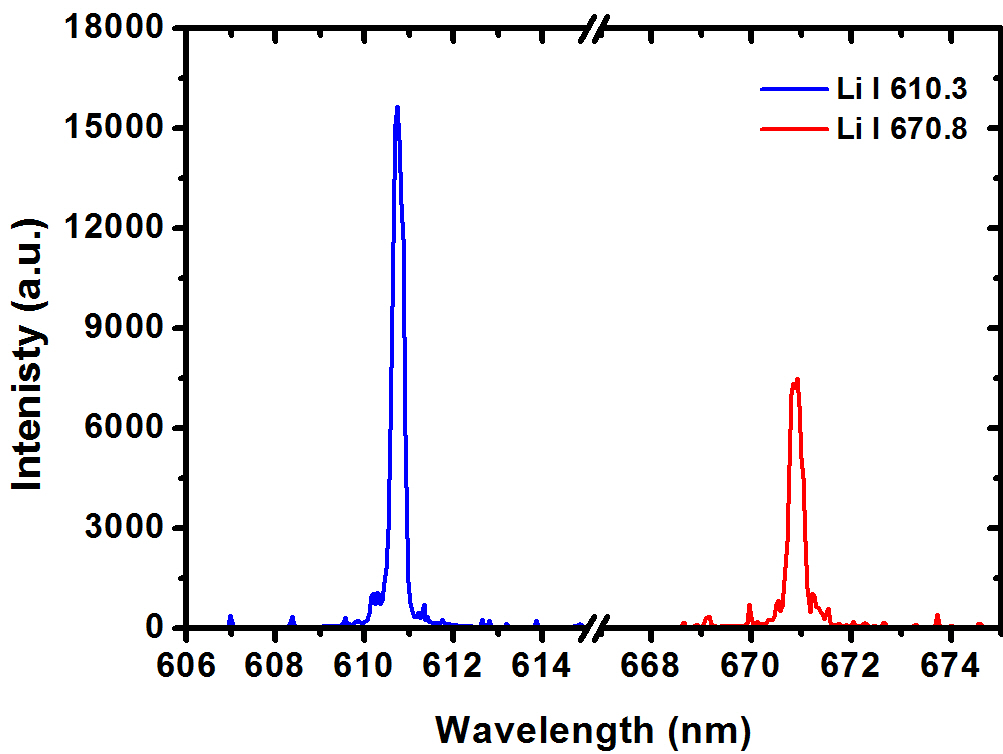}
\caption{\label{two_line} Emission profile in front ablation of Li I 610.3 nm and Li I 670.8 nm at 400 ns time delay in vacuum.}
\end{center}
\end{figure}

\subsection{Line spectra and plasma parameters:}

A typical energy level diagram showing Li I 610.3 nm (non-resonance) and Li I 670.8 (resonance) nm lines is shown in Fig. \ref{E_label} . Fig. \ref{two_line} shows typical spectrum lines Li I 670.8 nm and 610.3 nm in front ablation case at 2$\times10^{-6}$ mbar. As can be seen from the figure both the resonance (670.8 nm) and non-resonance (610.3 nm) lines are present with sufficient intensity under these conditions.

\begin{figure}
\begin{center}
\includegraphics[width = 6 cm]{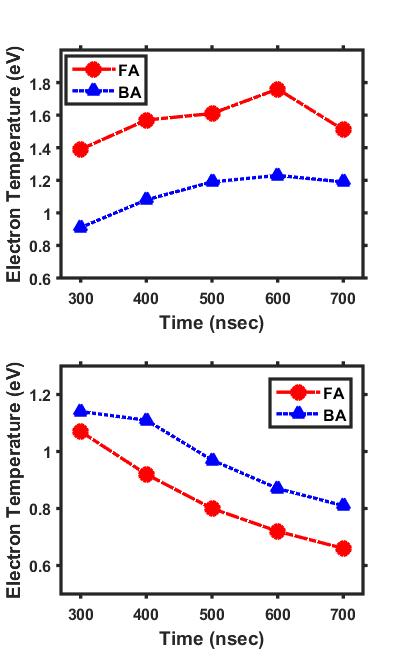}
\caption{\label{temp} Temporal evolution of electron temperature in FA and BA geometries in vacuum (top) and at 2$\times$10$^{-1}$ mbar (bottom).}
\end{center}
\end{figure}

The electron temperature has been calculated by using ratio of two neutral lithium line intensities i.e. Li I 610.3 nm and Li I 670.8 nm using the Boltzmann relation.\cite{smukasa2009temp}
Here, we would like to mention that one of the factors that can affect line intensities is the optical thickness of the plasma. As Li I 670.3 nm line originates from ground state, it is likely to be more prone to optical thickness. However, in the present experiment (as will be discussed latter) we rule out the contribution of optical thickness as neutral density is rather low as compared to electron/ion density  and the plasma thickness is very small (few mm).
\begin{equation}
\frac{I_{ij}}{I_{kl}}=\frac{\nu_{ij}A_{ij}g_{i}}{\nu_{kl}A_{kl}g_{k}} exp\frac{-(E_i - E_k)}{k_{\beta}T_e}
\end{equation}

Where I is the line intensity of the transition between two energy levels, $\nu$ is the frequency of the spectral line, A is Einstein’s coefficient, g represents the statistical weight of the energy level, E is the energy, $k_\beta$ is the Boltzmann constant, $T_e$ is the electron temperature, and the subscripts i, j, k and l represent the energy levels. The temporal evolution of electron temperature is shown in Fig. \ref{temp}. We would like to mention that obtained electron temperatures are less than the energy difference between the upper levels of these two transitions and hence the two line ratio method is applicable. The uncertainty in temperature measurements is around 5 - 10\%. As can be seen from Fig. \ref{temp} that under vacuum conditions electron temperature goes on increasing with time contrary to the case of 2$\times 10^{-1}$ mbar where opposite trend is noticed. Probably, the increase in temperature with time in vacuum may be due to inverse bremsstrahlung absorption, whereas the decrease at higher pressures may be due to increased collisions with the ambient which results in decrease in electron temperature.\cite{sivakumar2014optical}

\begin{figure}
\begin{center}
\includegraphics[width = 6 cm]{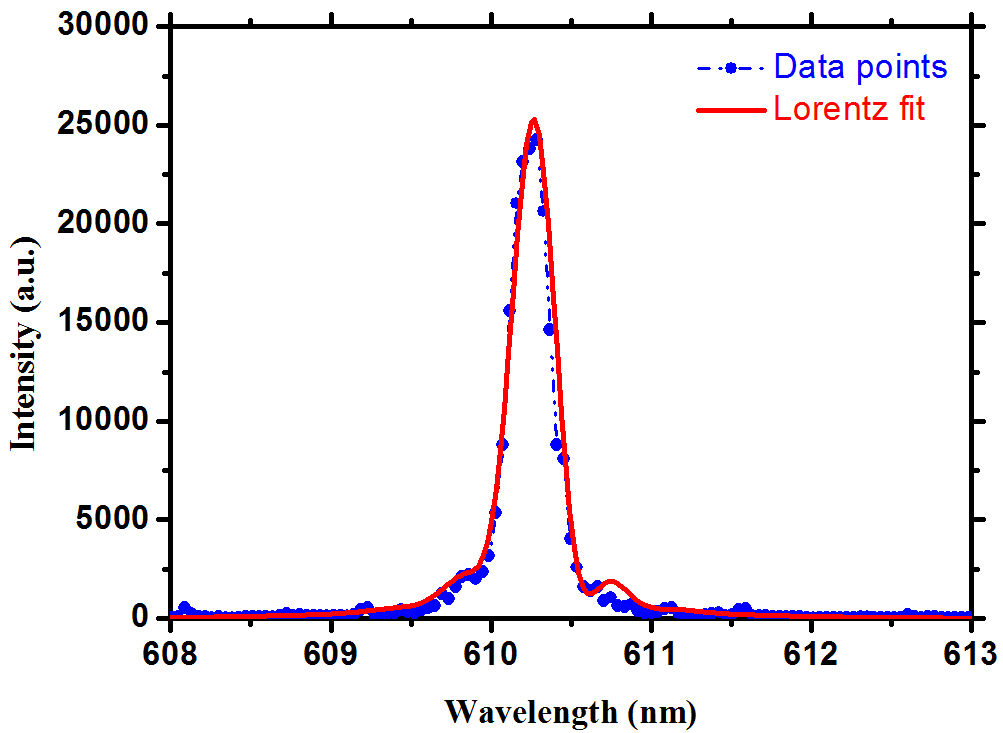}
\caption{\label{lorentfit} A typical Lorentzian fit at 300 ns in FA in vacuum.}
\end{center}
\end{figure}

\begin{figure}
\begin{center}
\includegraphics[width = 6 cm]{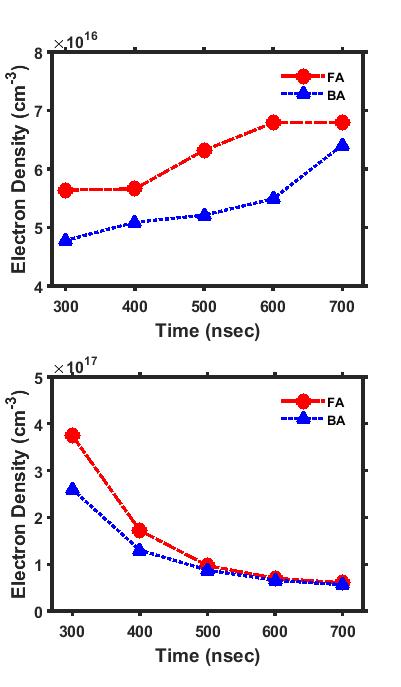}
\caption{\label{density} Electron densities in two geometries (i.e. FA and BA) and in vacuum (top) and 2$\times$10$^{-1}$ mbar pressures.}
\end{center}
\end{figure}

Electron density is determined from Stark broadening of 610.3 nm line by,\cite{bekefi1976principles, HRGriem}

\begin{equation}
\Delta \lambda_{\frac{1}{2}} = 2 \omega \frac{n_e}{10^{16}} \quad \AA
\end{equation}

where, $n_e$ is the electron density, $\Delta\lambda_\frac{1}{2}$ $\AA$ is the full width at half maxima (FWHM), $\omega$ is the electron-impact half width parameter which is weakly temperature dependent. $\omega$ is taken from reference \cite{HRGriem}. A typical Lorenzian fit for Li I 610.3 nm line is shown in  Fig. \ref{lorentfit}. Doppler and pressure broadening are neglected in the present case which have negligible contributions. Electron density shows slight increase in vacuum but shows decreasing trend at 2$\times 10^{-1}$ mbar (Fig. \ref{density} ). This can be anticipated that under vacuum conditions increase in temperature may result in more ionization and increase in density is expected.\cite{joshi2010effect} On the other hand recombination processes are likely to dominate at higher pressure and hence decrease in the density.\citep{joshi2010effect}

To ensure whether LTE condition is followed, we examined the validity of the McWhirter criteria for the present plasma condition which states that the density for LTE condition should follow the following relationship,\cite{bekefi1976principles}

\begin{equation}
n_e\geq 1.4\times10^{14}T_e^{\frac{1}{2}}(\Delta{E})^3\quad cm^{-3}
\end{equation}

Where $T_e$ (eV) is the electron temperature and $\Delta$E (eV) is the energy gap between two states of shorter wavelength Li I 610.3 nm. In our experiments for the maximum electron temperature of $\sim$1.76 eV and for the largest energy gap (2.031 eV) of the selected transition (610.3 nm) this criteria predicts a lower limit of $n_e$ = 2.95$\times$10$^{15}$ cm$^{-3}$. Estimated densities in the present case (Fig. \ref{density}) are always higher than this value and thus validate the LTE condition.

\subsection{Composition analysis}
\begin{figure}
\begin{center}
\includegraphics[width = 6 cm]{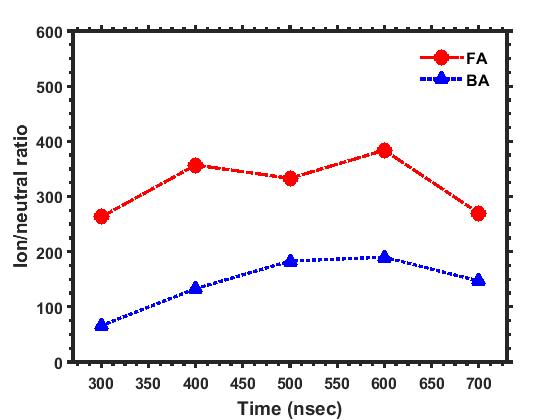}
\caption{\label{nist_6} Temporal evolution of ion-neutral ratio in front and back ablation geometry in vacuum using NIST LIBS database.\citep{nist}}
\end{center}
\end{figure}

\paragraph{(a) Ion/neutral (Ni/N)} Ion to neutral ratios for both FA and BA are estimated from NIST LIBS.\cite{nist} From NIST LIBS the ion/neutral abundance can be estimated (using Saha Equation) by entering electron temperature/density values. Fig. \ref{nist_6} represents ion to neutral ratio in case of FA as well as and BA geometries. In both the cases the ratio slightly increases with time and then deceases and has values greater than 1. This signifies that the population of ion is always higher in both the geometries as compared to neutrals. However, in case of front ablation this ratio is nearly about 2-4 times higher compared to back ablation plume. This reveals the higher abundance of ions in FA as compared to BA plumes. This can be expected as the electron temperature is much higher in case of FA (Fig. \ref{temp} ) leading to increased ionization.\citep{joshi2010effect}

\begin{figure}
\begin{center}
\includegraphics[width = 6 cm]{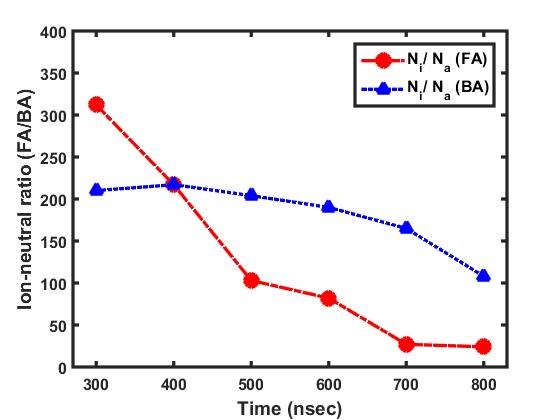}
\caption{\label{nist_1} Temporal evolution of ion-neutral ratio in front and back ablation geometry at 2$\times$10$^{-1}$ mbar using NIST LIBS database.\citep{nist}}
\end{center}
\end{figure}

\begin{figure}
\begin{center}
\includegraphics[width = 6 cm]{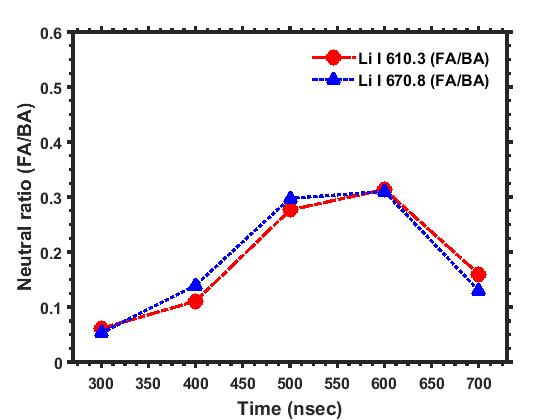}
\caption{\label{adas_pec_vac} Temporal evolution of neutral ratios ($\frac{Na_{FA}}{Na_{BA}}$) in two geometries calculated from ADAS (Eq. 5) using two different lines Li I 610.3 nm and Li I 670.8 nm in vacuum.}
\end{center}
\end{figure}

\begin{figure}
\begin{center}
\includegraphics[width = 6 cm]{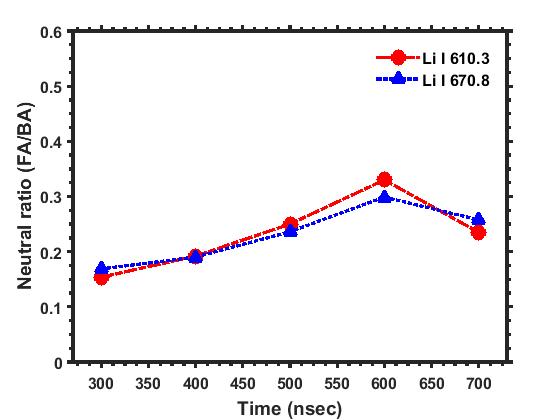}
\caption{\label{fig_part_fn1} Temporal evolution of neutral ratio ($\frac{Na_{FA}}{Na_{BA}}$) in two geometries by considering LTE condition (Eq. 6) for two lines Li I 610.3 nm and Li I 670.8 nm in vacuum.}
\end{center}
\end{figure}

\begin{figure}
\begin{center}
\includegraphics[width = 6 cm]{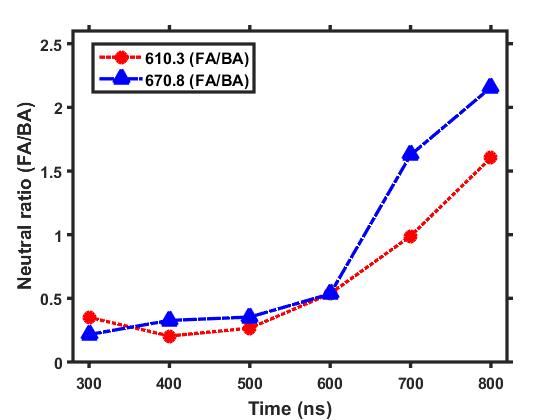}
\caption{\label{adas_1} Neutral ratio ($\frac{Na_{FA}}{Na_{BA}}$) calculated in two different geometries (i.e. FA and BA) using ADAS (Eq. 5) for two lines Li I 610.3 nm and Li I 670.8 nm at 2$\times$10$^{-1}$ mbar pressure.}
\end{center}
\end{figure}

For 2$\times$10$^{-1}$ mbar pressure again $n_i/n_a$ is also always higher than 1 which indicates at this pressure also ions dominate in the plasma plume. However, it decreases faster in case of FA as compared to BA. It can be noted that this decreasing trend with time is in confirmation with decrease in temperature which again can be explained by the decrease in ionization. As mentioned earlier, the overall behaviour can be attributed to combination of ionization recombination processes.

\paragraph{(b) Neutral abundance ($N_{FA}$/$N_{BA}$)}

We have adopted two approaches to calculate neutral abundance in these two geometries. In the first case we have used ADAS photon emissivity coefficients (PEC) to estimate the line intensity.\cite{adas} The intensity for a particular line can be given by the equation,\cite{joshi2010effect}

\begin{equation}
I = F_{exp} (N_e N_a PEC_{excit} + N_e N_i PEC_{recom})
\end{equation}

Where PEC$_{excit}$ is photon emissivity coefficient (photon $cm^{-3}$ $s^{-1}$) for electron impact excitation and PEC$_{recom}$ is photon emissivity coefficient for recombination (photon $cm^{-3}$ $s^{-1}$); $N_e$, $N_a$ and $N_i$ are electron, neutral and ion densities respectively and F$_{exp}$ is the experimental factor.

As F$_{exp}$ is same for both FA and BA, ratio of neutrals for forward to backward ablation ($N_{FA}/N_{BA}$) can be estimated by,

\begin{equation}
\frac{I_{FA}}{I_{BA}} = \frac{N_e N_{FA} PEC_{excit} + N_e N_i PEC_{recom}}{N'_e N'_{BA} PEC'_{excit} + N'_e N'_i PEC'_{recom}}
\end{equation}

where $N_e$, $N_{FA}$, $N_i$, PEC$_{exc}$, PEC$_{recom}$, $N'_e$, $N'_{BA}$, $N'_i$, PEC'$_{excit}$ and PEC'$_{recom}$ represent electron density, neutral density, ion density, photon emissivity coefficients for excitation and photon emissivity coefficient for recombination in FA and BA geometries respectively. $I_{FA}$ and $I_{BA}$ represents the line intensity in FA and BA respectively.  The values of ($N_{FA}$/$N'_{BA}$) for different times are shown in Fig. \ref{adas_pec_vac}. It can be seen that the ratios calculated from both the lines match well within experimental uncertainties as well as ADAS limitations. As in ADAS, PECs are self-consistent and consider the particulars levels involved in the transitions, an estimate of neutral population appears reasonably well.

In the second approach considering LTE condition (as discussed earlier), the line intensity can be given by the equation given below,\cite{ahamer2018}

\begin{equation}
I = \frac{F_{exp} g_u h\nu A_{ul} N exp^{-E_u/k_\beta T}}{Z(T)}
\end{equation}

Where, $F_{exp}$ is experimental factor, $g_u$ is statistical weight of upper level, h is Planck constant, $\nu$ is transition frequency, N is neutral density, $A_{ul}$ is transition probabilities, $E_u$ is upper level energy, $k_\beta$ is Boltzmann constant, T is plasma temperature, Z is partition function. From this equation the neutral abundance for FA and BA can be estimated by the following relationship,

\begin{equation}
\frac{N_a}{N'_a} = \frac{\frac{I_{FA}}{I_{BA}} \times \frac{Z_{FA}}{Z_{BA}}}{exp[- \frac{E_u}{k_\beta T_{FA}} + \frac{E_u}{k_\beta T_{BA}}]}
\end{equation}

Here, all the notations bear their standard meaning as mentioned earlier. We have calculated neutral ratios in FA and BA geometry using this relation and it is depicted in Fig. \ref{fig_part_fn1}. It can be seen that more neutrals are produced in case of back ablation.


Similar calculations, as mentioned above for 2$\times$10$^{-6}$ mbar, have also been done for 2$\times$10$^{-1}$ mbar pressure and are shown in the following figures. Fig. \ref{adas_1} and Fig. \ref{parfn_1mb} shows the neutral ratio calculated Eq. 5 and 6 respectively. In this case the values obtained using 610.3 nm and 670.8 nm lines also match within experimental uncertainties. Further, both the calculations show consistency up to 800 ns within the experimental limit for these two lines (Li I 610.3 and Li I 670.8 nm). These two figures show the neutrals are less at initial time approximately up to 700 ns in FA plasma plume in comparison to BA one although at later times (i.e. $>$ 700 ns) the neutral density is higher in FA compared to BA case. Again, it can be attributed to the higher ionic contribution at initial stage in case of FA. However, with evolution of time neutral population increases probably because of overall increase in recombination processes at this pressure. It can be noted that under vacuum conditions neutral density is higher in case of BA as compared to FA within our observation time window. This can also be understood  by the interplay of recombination/ionization processes.

\begin{figure}
\begin{center}
\includegraphics[width = 6 cm]{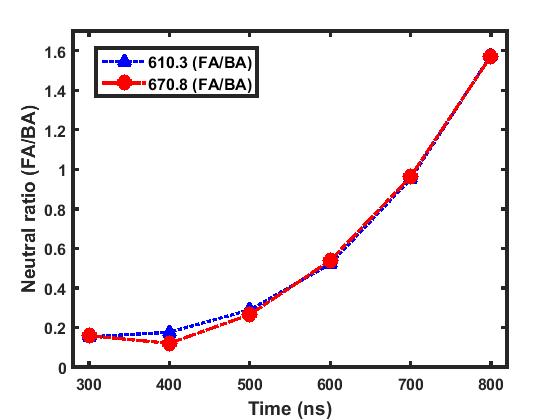}
\caption{\label{parfn_1mb} Neutral ratio in FA and BA by considering LTE condition (Eq. 6) at 2$\times$10$^{-1}$ mbar pressure.}
\end{center}
\end{figure}


Again we would like to emphasize that optically thin plasma conditions are met in the present experiment as neutral density is 1-2 orders smaller as compared to electron density. Further, this fact is also supported by the results where the neutral intensity ratios calculated from both 610.3 nm and 670.8 nm lines match well (Fig. \ref{adas_pec_vac} and Fig. \ref{parfn_1mb}) within experimental uncertainties.

Here it can be mentioned that evolution of neutral abundance for BA and FA geometries will depend on the respective ablation mechanisms as well evolution of electron density and temperature which in turn will affect ionization recombination processes in the generation of neutrals.

\section{Conclusion}

In this present report we have demonstrated compositional analysis of ions and neutrals in the evolving lithium plasma plume in FA and BA geometries in vacuum as well as 2$\times$10$^{-1}$ mbar pressure. Ion to neutral ratios is estimated from NIST (LIBS). It is found that ions dominate for both the cases. Further we considered two emission lines i.e. one resonance and other non-resonance and from their intensity ratios we have determined the neutral abundance in these two cases. In vacuum the neutral evolution with time shows that for FA they are always less as compared to BA. Analysis shows that neutrals gain with time in case of FA as time evolves at 2$\times$10$^{-1}$ mbar. In short in this work we have attempted and described an approach for the compositional analysis of evolving plasma in two different geometries. We believe that present work will have interesting implications for future works regarding evolution of compositional analysis in various scenarios of laser induced plasma e.g. thin film deposition, neutral beam generation etc.

\section*{References}

\bibliography{mybibfile}
\end{document}